\documentclass[journal]{IEEEtran}

\usepackage{amsmath,amssymb}
\usepackage{booktabs}
\usepackage{array}
\usepackage{multirow}
\usepackage{graphicx}
\usepackage{graphicx}
\usepackage{tikz}
\usetikzlibrary{arrows.meta,positioning,shapes.geometric,fit}

\usepackage{cite}
\usepackage[hidelinks]{hyperref}
\usepackage{orcidlink}
\usepackage[nameinlink,noabbrev]{cleveref}

\title{System One Models for Wireless Decision-Making: Applications and Performance Evaluation}

\author{
Masoud Rahimi \orcidlink{0009-0007-6374-2756},
S. M. Matin Alemohammad \orcidlink{0009-0003-3636-3460},
Hamid Behroozi \orcidlink{0000-0001-9294-3134},~\IEEEmembership{Member,~IEEE},
and Mahdi Nouri,~\IEEEmembership{Senior Member,~IEEE}
\thanks{
The authors are with the Department of Electrical Engineering, Sharif
University of Technology, Tehran 1458889694, Iran
(e-mail: masoud.rahimi@sharif.edu; matin.alemohammad79@sharif.edu).
}
}
\begin{document}

\maketitle

\begin{abstract}
Many wireless control tasks require repeated selection of a single action from a finite feasible set under stringent latency and reliability constraints. While large language models (LLMs) have recently emerged as general-purpose decision engines, their autoregressive generation mechanism is not naturally aligned with such bounded control problems. This paper investigates System-One models, which directly learn probability distributions over explicitly defined decision spaces, as a lightweight alternative for wireless decision-making. We formalize their decision structure and learning objective, identify their applicability across physical-layer control, radio resource management, mobility, network slicing, and network operations, and evaluate their practical behavior through representative wireless case studies. Using Jev as a System-One implementation, we benchmark decision quality and client-observed latency against generative LLMs and conventional baselines. In receive-antenna selection, Jev delivers up to an 8.5x reduction in median response latency relative to the evaluated LLMs, although this gain comes with a loss in decision quality compared with stronger task-specific alternatives. More notably, in intent-conditioned RAN slicing, Jev achieves utility comparable to the evaluated LLMs while providing more than a 3.5x reduction in median response latency. Complementary evidence from edge-service orchestration further shows that faster decisions do not necessarily translate into lower end-to-end service latency. These results expose a fundamental quality/latency tradeoff and position System-One models not as replacements for numerical optimization, but as a promising decision interface for latency-sensitive, bounded, and intent-driven wireless control.
\end{abstract}


% =============================================================================
\section{Introduction}
\label{sec:introduction}
% =============================================================================

Artificial intelligence is becoming an integral component of future wireless
networks, spanning physical-layer processing, radio-resource management,
mobility, network operations, and service orchestration
\cite{itu_m2160,airan2024vision}. Many of these functions differ fundamentally
from open-ended generative tasks: they repeatedly map an observed network state
to a bounded operational decision, such as selecting a modulation-and-coding
scheme (MCS), beam, serving cell, scheduling policy, antenna subset, or
remediation action. Such decisions must often be produced under tight latency,
reliability, and feasibility requirements.

Wireless intelligence has traditionally addressed these problems through
model-based optimization and task-specific learning. Optimization provides
explicit objectives and constraints but can require iterative numerical
procedures whose online cost grows with problem complexity. Learned policies can
amortize this computation and provide fast inference, but they are commonly
designed for a fixed task, state representation, and action space
\cite{oshea2017introduction,ye2018power,wang2018multichannel,feriani2021single}.
Foundation models offer a different objective: reuse learned representations and
knowledge across multiple tasks rather than training an independent model for
every function \cite{vaswani2017attention,bommasani2021opportunities}. Recent
work has explored this direction through telecom-specific language models
\cite{zou2024telecomgpt}, LLM adaptation for networking tasks
\cite{wu2024netllm}, and wireless-native foundation models operating on RF and
channel data
\cite{aboulfotouh2025radiofm,lwm2025,zou2026rfgpt,liang2026foundation}.

However, the appropriate role of an LLM depends on the structure of the task.
Natural-language intent interpretation, knowledge retrieval, planning, and
troubleshooting can benefit directly from generative language models. In
contrast, many radio and network-control tasks ultimately require a decision
from an explicitly admissible set. Autoregressive generation is then an
indirect interface: an answer is constructed token by token and subsequently
parsed, validated, and mapped to an executable action. This distinction has
already appeared in networking research. NetLLM, for example, uses
task-specific networking heads that bypass free-form generation and directly
produce valid task outputs \cite{wu2024netllm}. These observations motivate a
broader question: when the desired output is a bounded decision rather than
generated content, should decision inference itself be the native model
interface.

Recently, the term \emph{System-One model} has been introduced for
decision-oriented models that replace free-form sequence generation with typed
probabilistic queries over a declared answer space
\cite{typesafe2026systemone,laya2026}. At the interface level, such a model
receives an observed state $\mathbf{s}$, a decision query $q$, and an admissible
answer set $\mathcal{A}$, and returns a probability distribution
\begin{equation}
    \label{eq: conditional}
    p_\theta(Y \mid \mathbf{s}, q, \mathcal{A}),
    \qquad
    Y \in \mathcal{A},
\end{equation}
rather than a free-form token sequence. 
The distinction is not merely that the output is categorical. The query and
admissible outcomes are themselves conditioning variables, enabling a shared
model, in principle, to address different decision schemas without a separately
trained output head for every task. The interface also guarantees schema-valid
outputs and exposes probabilities that can support confidence-aware execution,
abstention, or escalation. The first public implementation, Jev, is reported by
TypeSafe to use a new model architecture, a parallel sampler, and a
calibration-oriented training method termed Reinforcement Learning for
Calibrated Decisions (RLCD) \cite{typesafe2026systemone}. By eliminating
token-by-token output generation, it is explicitly designed for low-latency
machine decision making; TypeSafe reports 70--500~ms end-to-end response times
and 40--200$\times$ speedups over frontier LLMs on its System-One-oriented
workloads. These figures are vendor-reported and workload dependent, but they
motivate evaluating the computational advantage directly in wireless settings.

AnyJev provides a complementary open-source route by converting pretrained
LLMs into System-One-style typed decision interfaces without fine-tuning their
backbone weights. It obtains bounded decisions through non-generative readouts
and supports zero-label debiasing as well as lightweight, per-question
calibration or closed-form decision heads \cite{anyjev2026}. 
Thus, the System-One interface can be separated from the architecture used to implement it.

\begin{figure}[t]
    \centering
    \includegraphics[width=\columnwidth]{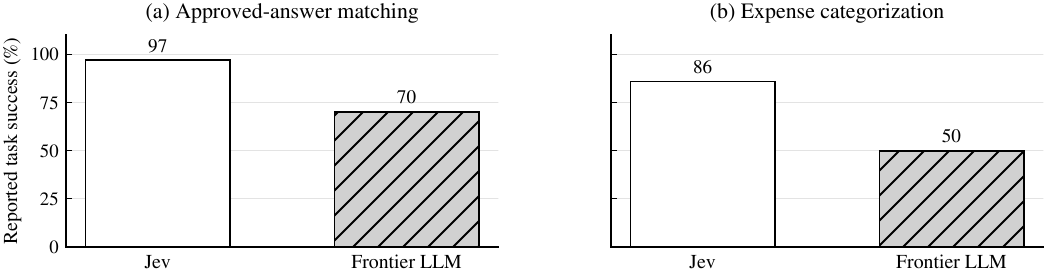}
    \caption{Reported quality on two production decision tasks evaluated by
    Deel. The first panel reports approved-answer matching accuracy; the second
    reports expense-categorization agreement with human reviewers. Results were
    published by TypeSafe AI \cite{typesafe2026deel}.}
    \label{fig:deel-reported-quality}
\end{figure}

Preliminary third-party evidence has also been reported from a customer-side
evaluation by Deel. On two production decision tasks, the published results
show approved-answer matching increasing from 70\% with a frontier LLM to 97\%
with Jev, and expense-categorization agreement with human reviewers increasing
from 50\% to 86\%, as summarized in Fig.~\ref{fig:deel-reported-quality}
\cite{typesafe2026deel}. Because the evaluated datasets, sample sizes,
comparator models, and prompting configurations have not been publicly
documented, these figures should be interpreted as customer-reported evidence
rather than an independently reproducible benchmark.

Wireless networks provide a natural testbed for this model class because many
control functions combine repeated state-conditioned inference with bounded
actions. Candidate examples include link adaptation, beam and antenna
selection, handover, slice admission, scheduling-policy selection,
energy-control modes, and fault-response routing. The fit is not universal.
High-dimensional continuous outputs such as channel estimates, arbitrary
precoding matrices, or waveforms are better handled by specialized estimators,
regression or generative models, or optimization unless they are decomposed
into bounded sub-decisions. Likewise, hard physical and protocol constraints
should remain under deterministic or optimization-based control rather than
being delegated solely to a learned probabilistic model.

This perspective also motivates a hybrid System-One/System-Two architecture
for wireless intelligence. Routine decisions with sufficiently high confidence
can follow a fast decision path, whereas ambiguous, out-of-distribution, or
high-risk states can trigger additional measurements or be escalated to an
optimizer, simulator, or LLM-based reasoning agent. The resulting question is
therefore not whether System-One models should replace existing wireless
algorithms, but where a shared decision-native model provides measurable value
in decision quality, latency, calibration, robustness, and cross-task reuse,
and how it should interact with established model-based and learning-based
methods.

The main contributions of this article are as follows:
\begin{itemize}
    \item We formalize System-One models as schema-conditioned probabilistic
    decision models and clarify their relationship to conventional classifiers,
    task-specific learned policies, and autoregressive LLMs.

    \item We identify the structural properties that make a wireless task
    suitable for this interface and organize representative opportunities
    across PHY, MAC/RRM, mobility, and network management.
    % , while delineating
    % tasks for which continuous estimation, optimization, or generative
    % reasoning remains more appropriate.

    \item We evaluate the decision-model paradigm through receive-antenna
    selection and intent-conditioned RAN slicing, and complement these
    experiments with independently reported intent-driven edge-service
    orchestration results. The comparison considers conventional baselines and
    generative LLMs and examines decision quality, latency, and end-to-end
    utility.
\end{itemize}

The remainder of the article is organized as follows.
\Cref{sec:s1-structure} develops the System-One decision interface and learning
formulation. \Cref{sec:wireless-opportunities} examines candidate wireless
applications and suitability criteria.
\Cref{sec:experimental-evaluation} presents the evaluation methodology and case
studies, while \cref{sec:challenges} discusses open problems in wireless-native
decision models, calibration, safety, and deployment.

% =============================================================================
\section{System-One Models: Functional Structure and Learning Objectives}
\label{sec:s1-structure}

As introduced in \cref{eq: conditional}, given an observed state $s$, a decision query $q$, and a finite admissible answer space $\mathcal{A}$, the model provides a conditional distribution 
$p_{\theta}(Y\mid s,q,\mathcal{A})$ over the admissible outcomes. 
Here, $s$ contains the information available for decision making, $q$ specifies the decision criterion, and $\mathcal{A}$ defines the allowable outputs.

The primary model output is therefore a probability distribution over decisions rather than a free-form sequence of generated tokens.

This formulation defines the model at the level of its probabilistic interface and task structure. It does not depend on a particular neural backbone, prediction head, or optimization algorithm. Different implementations may realize the same conditional decision distribution using different architectures and training procedures.

The answer space can encode categorical, binary, or ordered decisions. For a categorical decision with
$\mathcal{A}=\{a_1,\ldots,a_K\}$, the predicted answer is
\begin{equation}
 \hat a
 =\underset{a_k\in\mathcal{A}}{\arg\max}\;
 p_{\theta}(Y=a_k\mid s,q,\mathcal{A}).
\end{equation}

Binary decisions are simply the special case $K=2$. The same formulation also covers ordered decision spaces. If each outcome $a_k$ is associated with an ordered numerical level $v_k$, let $V$ denote the corresponding discrete-valued random variable, i.e., $V=v_k$ when $Y=a_k$. The predictive distribution then induces the scalar summary
\begin{equation}
 \hat r = \mathbb{E}[V\mid s,q,\mathcal{A}].
\end{equation}
This quantity is the conditional expectation over the explicitly declared numerical levels. It therefore remains tied to the specified decision schema and should not be interpreted as an unrestricted continuous regression output.

Learning is formulated as conditional distribution learning over decision spaces. Let
\begin{equation}
    x=(\mathbf{s},q,\mathcal{A}),
\end{equation}
and let $Y$ denote the target decision outcome. The learning problem is
\begin{equation}
    \theta^\star
    =
    \arg\min_\theta
    \mathbb{E}_{(x,Y)\sim\mathcal{D}}
    \left[
        \ell\!\left(
            p_\theta(\cdot\mid x),Y
        \right)
    \right],
\end{equation}
where $\ell$ evaluates the quality of the reported predictive distribution.

Strictly proper scoring rules provide a natural objective for this setting because their expected value is optimized when the reported distribution matches the underlying conditional distribution. Cross-entropy and the Brier score are standard examples. For instance, cross-entropy with a target distribution $\mathbf{t}$ is
\begin{equation}
    \mathcal{L}_{\mathrm{CE}}
    =
    -\mathbb{E}_{(x,\mathbf{t})\sim\mathcal{D}}
    \left[
        \sum_{k=1}^{K}
        t_k
        \log p_\theta(Y=k\mid x)
    \right],
\end{equation}
with hard-label training obtained as the special case
$t_k=\mathbf{1}[k=Y]$.

The probabilistic learning objective is distinct from the optimization procedure used to fit the model. Supervised likelihood-based training, calibration-oriented objectives, reinforcement learning, and hybrid procedures provide different mechanisms for learning the conditional distribution
$p_\theta(Y\mid\mathbf{s},q,\mathcal{A})$
\cite{typesafe2026systemone,laya2026}. These procedures therefore represent alternative learning strategies rather than defining characteristics of the System-One model class.

The distinction from a conventional fixed-label classifier lies primarily in how the prediction problem is specified. A conventional classifier typically represents
\begin{equation}
    p_\theta(Y\mid\mathbf{s}),
\end{equation}
where the task and label space are largely fixed during model development. 
In the System-One abstraction, the query $q$ and admissible answer space $\mathcal{A}$ are explicit parts of the model input.
The same pretrained model can therefore perform
different tasks without task-specific fine-tuning: changing $q$ and
$\mathcal{A}$ changes the decision problem and its output space while keeping
$\theta$ fixed. Unlike a conventional fixed-label classifier, it does not
require a newly trained output head for each task. This cross-task reuse is
a central capability, although accuracy and calibration on each new task
must still be evaluated.

The distinction from an autoregressive language model is equally fundamental at the level of output factorization. An autoregressive language model represents a distribution over token sequences,
\begin{equation}
    p_\psi(w_{1:L}\mid X)
    =
    \prod_{\ell=1}^{L}
    p_\psi(w_\ell\mid X,w_{<\ell}),
\end{equation}
and constructs its output through sequential token prediction. 
A System-One interface, in contrast, exposes a probability distribution directly over the declared decision space and uses that distribution to select or score admissible actions.
Autoregressive language models can also perform classification and structured prediction, but in that case the decision is expressed through, or derived from, a token-generation process rather than being the native modeled output space.

Finally, probabilistic output must be distinguished from calibration. Normalization guarantees only that the reported values form a valid probability distribution; it does not guarantee that those probabilities correspond to empirical frequencies of correctness. A model can therefore produce normalized probabilities while remaining systematically over-confident or under-confident. Calibration is an empirical property of the learned predictor relative to a data distribution and deployment regime, and it must be assessed independently of the structural definition of a System-One model.

% =============================================================================
\section{System-One Decision Opportunities in Wireless Networks}
\label{sec:wireless-opportunities}

% =============================================================================

Having established the probabilistic formulation of System-One models, we now
examine where this decision interface may be useful in wireless networks. The
opportunities span PHY-level control, MAC and radio resource management,
mobility, and network operations. We first identify the characteristics of a
suitable decision task and then use these criteria to organize the candidate
applications and select representative case studies.

\subsection{Characteristics of Suitable Wireless Decisions}
\label{subsec:suitable-decisions}

Not every wireless function is an appropriate target for a System-One model.
Suitability is determined less by the protocol layer than by the structure,
timing, and risk profile of the decision. The following characteristics provide
a practical screening framework.

\emph{1) A bounded and explicit decision space:}
The task should admit a finite set of meaningful alternatives, such as an MCS
index, beam identifier, serving cell, scheduling mode, alarm class, or
sleep/wake command. This structure allows the model to return a normalized
distribution over feasible outcomes and enables the controller to reject
out-of-schema actions by construction. Tasks whose native output is a
high-dimensional continuous object, such as an unconstrained precoding matrix
or waveform, are less direct candidates unless the decision is discretized,
factorized into several bounded queries, or combined with a downstream
optimizer.

\emph{2) Repeated state-conditioned decisions:}
The same decision semantics should recur across many network states. Wireless
control commonly has this form: channel, traffic, queue, mobility, and load
measurements are observed repeatedly, and an action is selected from a
well-defined set \cite{feriani2021single}. This favors amortized inference, in
which computational effort is shifted to training and decisions are produced
quickly online. In contrast, one-off engineering tasks that require open-ended
design, extensive explanation, or long-horizon decomposition are better suited
to optimization tools, simulators, or deliberative agents. A System-One model
may still support such workflows, but only as a bounded decision component.

\emph{3) A compatible end-to-end control timescale:}
The relevant requirement is not model latency alone, but the complete delay
from state acquisition through preprocessing, inference, validation, signaling,
and actuation. O-RAN identifies near-real-time control loops operating roughly
between \(10\)~ms and \(1\)~s and non-real-time loops above \(1\)~s, while
sub-\(10\)-ms loops remain closely tied to embedded RAN processing and hardware
constraints \cite{oran2024dapp}. System-One models are therefore initially more
plausible for decisions with millisecond-to-second budgets than for
symbol-level processing. Tighter loops remain possible only if the complete
implementation, including the input interface and safety checks, satisfies the
deadline.

\emph{4) Uncertainty that changes the downstream action:}
Probabilistic output is most valuable when confidence determines whether a
decision is executed, verified, deferred, or passed to a fallback method.
Uncertainty-aware interference management, for example, has been used to
switch between a fast learned solution and a conventional algorithm when the
learned prediction is insufficiently trustworthy
\cite{lee2025selfimproving}. A suitable System-One task should similarly define
what the controller does with low confidence; otherwise, returning a
probability distribution provides little operational benefit. This requirement
also makes calibration under deployment shift as important as average
accuracy.

\emph{5) Reusable structure across related decision schemas:}
The strongest motivation for a shared decision model arises when states and
decision semantics vary across cells, configurations, or objectives, but retain
common structure. Conventional DRL policies may require substantial retraining
when state or action spaces change, reducing sample efficiency and limiting
generalization \cite{zhang2025decisiontransformer}. A System-One model is
particularly relevant when the answer set or decision query must be specified
at runtime. However, schema flexibility should not be assumed to imply
zero-shot competence: transfer to new queries, environments, and action sets
must be measured explicitly.

\emph{6) Enforceable constraints and a safe fallback:}
Many wireless decisions are subject to hard power, spectrum, reliability,
fairness, and protocol constraints. A probabilistic model should not be solely
responsible for enforcing requirements that can be checked deterministically.
The selected action should pass through a feasibility layer, and the system
should retain an optimization-based, rule-based, or human fallback for
out-of-distribution or high-risk states. This hybrid view is consistent with
recent knowledge-driven wireless AI frameworks, which combine learned
components with physical models and optimization structure
\cite{sun2025knowledge}.

No single characteristic is sufficient by itself. Suitability increases when a
task combines a bounded action space, repeated state-conditioned execution, an
achievable end-to-end deadline, and a meaningful use for calibrated
confidence, while also permitting explicit constraint checking and a credible
fallback. These criteria are applied across the protocol stack in the following
subsections and used to select the representative case studies in
\cref{sec:experimental-evaluation}.

\subsection{PHY-Level Decisions}
\label{subsec:phy-decisions}

At the PHY layer, the most direct opportunities are not arbitrary signal
reconstruction problems, but discrete configuration decisions made from
channel measurements, feedback, and recent transmission outcomes. Link
adaptation is a representative example. In NR, the transmitter selects an MCS
from standardized tables that jointly determine the modulation order and target
code rate \cite{3gpp38214}. Let
\(\mathbf{s}^{\mathrm{LA}}_t\) collect features such as effective SINR, CQI,
HARQ acknowledgements, rank and precoding indicators, Doppler or mobility
information, and recent BLER. With the admissible MCS set \(\mathcal{M}\), a
simplified reliability-constrained decision is
\begin{equation}
    \begin{aligned}
        m_t^\star
        ={}&\arg\max_{m\in\mathcal{M}}
        r_m\!\left[1-\widehat P_{\mathrm{BLER}}
        (m\mid\mathbf{s}^{\mathrm{LA}}_t)\right]\\
        \text{s.t.}\quad&
        \widehat P_{\mathrm{BLER}}
        (m\mid\mathbf{s}^{\mathrm{LA}}_t)\leq\epsilon ,
    \end{aligned}
    \label{eq:link-adaptation}
\end{equation}
where \(r_m\) is the spectral efficiency of MCS \(m\) and \(\epsilon\) is the
target BLER. Existing work has already shown that learning can complement
outer-loop link adaptation when the mapping from compact feedback to BLER is
difficult to model accurately \cite{martinvega2021link}. A System-One model
would return a distribution over \(\mathcal{M}\), rather than only the
maximizing index. This distribution could support conservative fallback to a
lower MCS, top-\(k\) verification, or abstention when the state is far from the
training distribution. The principal challenge is timing: the complete
feature-construction, inference, and scheduling path must fit within the link
adaptation deadline.

Beam selection has a similar bounded structure when analog or hybrid
beamforming uses a finite codebook
\(\mathcal{B}=\{b_1,\ldots,b_{K_B}\}\). The state may contain partial CSI,
received powers, position, angle estimates, blockage indicators, or previous
beam measurements, and the action is a beam or beam pair. A natural target is
\begin{equation}
    b_t^\star
    =\arg\max_{b\in\mathcal{B}}
    \widehat R(b\mid\mathbf{s}^{\mathrm{B}}_t),
    \label{eq:beam-selection}
\end{equation}
where \(\widehat R\) denotes predicted rate or received power. Recent
probability-map approaches explicitly predict the probability that each
candidate is optimal and adapt the number of beams retained for subsequent
training \cite{shao2025beam}. This is closely aligned with a typed decision
distribution: the controller may select one beam when confidence is high or a
small candidate subset whose cumulative probability exceeds a threshold when
confidence is diffuse.

Other bounded PHY decisions include antenna or spatial-layer selection,
receiver-mode selection, numerology or bandwidth-part selection, and choosing
among predefined detection, decoding, or channel-estimation procedures.
However, raw channel estimation, waveform generation, and unconstrained
precoder design produce continuous high-dimensional objects and are not direct
System-One tasks. They require either a specialized estimator, a generative or
regression model, or a decomposition in which the System-One model selects a
mode, codebook entry, or solver configuration while another component computes
the continuous parameters.

\subsection{MAC, RRM, and Mobility Decisions}
\label{subsec:rrm-mobility-decisions}

MAC and radio resource management expose a large number of repeated decisions,
but their suitability depends strongly on how the action space is represented.
Let 
\(\mathbf{s}_t\) 
include per-user channel indicators, queue
lengths, packet deadlines, slice requirements, interference, and cell load.
Many resource-management objectives can be expressed abstractly as
\begin{equation}
    \begin{aligned}
        a_t^\star
        ={}&\arg\max_{a\in\mathcal{A}(\mathbf{s}_t)}
        % \sum_{u} U_u\!\left(R_u(\mathbf{s}_t,a)\right)\\
        \sum_{u} U_u\!\left(\mathbf{s}_t,a\right)\\
        \text{s.t.}\quad&
        g_i(\mathbf{s}_t,a)\leq 0,\quad i=1,\ldots,I ,
    \end{aligned}
    \label{eq:rrm}
\end{equation}
where \(U_u\) captures rate, delay, or fairness utility and \(g_i\) represents
resource and QoS constraints. Directly treating every joint assignment of
users, resource blocks, beams, and power levels as one categorical action is
usually infeasible because the number of alternatives grows combinatorially.
A more credible System-One formulation is hierarchical: select a scheduling
rule, a user group, a slice-allocation profile, or a short list of candidate
schedules, and then use a constrained scheduler or optimizer to produce the
final allocation.

RAN slicing illustrates this intermediate level. The state contains slice and
UE metrics, while the action may select among scheduler policies or bounded
resource-sharing profiles intended to satisfy service-level intents.
Intent-aware learning has been evaluated for allocating scarce radio resources
among slices with heterogeneous requirements
\cite{nahum2024intent}. A typed decision model could expose the probability of
each scheduling profile and abstain when none is predicted to satisfy the
declared intents. Continuous power control can be handled similarly by using a
finite power codebook or by selecting a power-control policy; unrestricted
per-resource-block power vectors should remain with a continuous controller.

Mobility management provides an especially natural decision schema. For a UE
with serving cell \(c_0\) and candidate neighbors
\(\mathcal{C}_t=\{c_1,\ldots,c_K\}\), the action set is
\begin{equation}
    \mathcal{A}^{\mathrm{HO}}_t
    =\{\text{stay at }c_0,\text{ hand over to }c_1,\ldots,c_K\}.
\end{equation}
The state may combine serving and neighboring RSRP/RSRQ/SINR histories, UE
velocity, cell load, QoS, and recent handovers. A useful decision objective is
\begin{equation}
    J(a)=
    \mathbb{E}\!\left[
    \bar R(a)-\lambda_{\mathrm{f}}F(a)
    -\lambda_{\mathrm{p}}P(a)-\lambda_{\mathrm{o}}O(a)
    \right],
    \label{eq:handover-utility}
\end{equation}
where \(F\), \(P\), and \(O\) denote handover failure, ping-pong, and outage
costs. O-RAN traffic-steering work has demonstrated user-specific serving-cell
selection using near-real-time data and learned control policies
\cite{lacava2024traffic}. A System-One distribution over the current candidate
cells would additionally make confidence explicit: high-confidence actions may
be executed, whereas ambiguous cases can retain the serving cell, trigger
additional measurements, or invoke a slower policy.

\subsection{Network Management Decisions}
\label{subsec:network-management-decisions}

Network-management decisions operate on slower timescales and broader state
descriptions than PHY or MAC control. The state may include KPI windows, alarm
sequences, configuration changes, topology, traffic forecasts, security
events, and service-level objectives. If a set of approved remediation actions
is available, the problem can be represented as
\begin{equation}
    \begin{aligned}
        a_t^\star
        =\arg\min_{a\in\mathcal{A}^{\mathrm{NM}}}
        \mathbb{E}\!\big[&C_{\mathrm{svc}}(a)
        +\lambda_{\mathrm{e}}C_{\mathrm{energy}}(a)\\
        &+\lambda_{\mathrm{r}}C_{\mathrm{risk}}(a)
        \mid\mathbf{s}^{\mathrm{NM}}_t\big],
    \end{aligned}
    \label{eq:network-management}
\end{equation}
subject to coverage, capacity, security, and policy constraints. Possible
actions include monitor, collect additional evidence, restart a component,
rollback a configuration, throttle or isolate an entity, activate a predefined
xApp/rApp, change an energy-saving mode, or escalate to an operator.

Fault and anomaly response is a direct candidate when detection is followed by
a bounded playbook. O-RAN testbeds have demonstrated radio-originating anomaly
detection and mitigation through specialized xApps
\cite{bogucka2025oran}. A System-One model could jointly estimate an incident
class, severity level, and probability distribution over approved mitigation
actions. Low confidence should lead to observation or escalation rather than
an irreversible response. The same formulation applies to alarm
prioritization and root-cause ranking, although free-form root-cause
explanation is better delegated to an LLM or knowledge system.

Other opportunities include slice admission
\(\{\text{accept},\text{reject},\text{defer}\}\), selection among deployment
templates, traffic-steering activation, and cell or carrier sleep/wake control.
These actions are bounded, but their effects are coupled across space and time.
For example, switching off a lightly loaded cell may save energy while
violating coverage or overloading a neighbor. A System-One model should
therefore recommend a candidate management action and probability, while a
digital twin, constraint checker, or optimizer validates its predicted impact.

Natural-language intent interpretation is a less direct opportunity. User and
operator intents may be ambiguous, cross-domain, and require clarification or
decomposition; such tasks favor LLM-based agents. After an intent has been
translated into explicit service objectives and admissible configurations,
however, System-One models may support bounded choices such as policy
selection, feasibility classification, conflict severity, and remediation
routing. The boundary is therefore between open-ended intent understanding and
typed operational decision making, not simply between management and control.

\begin{table*}[t]
    \centering
    \caption{Representative bounded decision opportunities across the wireless stack.}
    \label{tab:wireless-opportunities}
    \renewcommand{\arraystretch}{1.12}
    \begin{tabular}{@{}p{0.12\textwidth}p{0.23\textwidth}p{0.19\textwidth}p{0.11\textwidth}p{0.21\textwidth}@{}}
        \toprule
        Decision & Representative state & Admissible output & Typical timescale & Principal System-One value \\
        \midrule
        Link adaptation & SINR/CQI, HARQ history, rank, mobility & MCS or link-configuration index & Slot to tens of ms & Reliability-aware distribution and conservative fallback \\
        Beam selection & Partial CSI, position, angles, past beams & Beam or top-\(k\) beam subset & Beam-coherence dependent & Adaptive candidate-set size from confidence \\
        Scheduling/RRM & Channels, queues, deadlines, slice KPIs & Policy, user group, or candidate schedule & Slot to near-RT & Rapid choice among constrained profiles \\
        Handover & Signal histories, velocity, load, QoS & Stay or target cell & Tens of ms to seconds & Explicit ambiguity and safe defer/measure action \\
        Slice admission & Resource state, SLA, predicted demand & Accept, reject, defer, or template & Seconds and above & Typed decision with feasibility screening \\
        Anomaly response & KPI/alarm history, topology, security events & Monitor, isolate, throttle, rollback, escalate & Near-RT to management & Risk-aware mitigation and abstention \\
        Energy control & Traffic forecast, coverage, neighbor load & Active/sleep mode or policy & Seconds to minutes & Candidate policy selection before validation \\
        \bottomrule
    \end{tabular}
\end{table*}

% =============================================================================
\section{Experimental Evaluation}
\label{sec:experimental-evaluation}

We examine three complementary scenarios: receive-antenna selection,
intent-conditioned RAN slicing, and intent-driven edge-service
orchestration. The first two are implemented in our simulation
framework, whereas the third draws on the independent study
in~\cite{li2026jev}. Together, they cover numerical selection,
joint semantic and numerical decision making, and integration with
network-service execution.

\subsection{Evaluation Methodology}

Our evaluation compares Jev with frontier LLMs and task-specific classical
baselines. Each method is evaluated on the same instances within
each task, and its selected action is scored by a common numerical
evaluator. Jev uses a categorical decision interface, while the LLMs
are instructed to return a single action identifier in JSON format.
Optimal actions and evaluator-generated scores are withheld from
model inputs.

The evaluation considers decision quality and response latency.
For API-based methods, latency is measured at the client for the
successful request, excluding deliberate pacing and retry waits.
It therefore reflects the serving infrastructure and communication
path as well as model execution. Classical methods are timed locally;
these measurements provide implementation-level references rather
than hardware-matched inference comparisons.

\subsection{Scenario I: Receive-Antenna Selection}

The first scenario considers a MIMO link with two transmit antennas
and eight candidate receive antennas, of which two are selected to
maximize instantaneous capacity. Ten independent Rayleigh-fading
channel realizations are evaluated at SNRs of 0, 10, and 20~dB,
yielding 30 channel--SNR instances and 28 admissible antenna pairs
per instance.

Exhaustive search provides the optimal reference. Practical baselines
include norm-based selection, a correlation-based method (CBM), and
random selection \cite{molisch2005capacity}. The AI methods receive the complex channel matrix,
SNR, and candidate actions. Input variants examine the effect of
capacity-oriented guidance and derived channel-strength and
correlation features. For the LLMs, the guided variant includes both
guidance and these features; Jev additionally separates guidance-only
and feature-assisted inputs.

Performance is assessed through relative capacity loss against
exhaustive search, optimal-action selection rate, and decision
latency. This scenario tests whether a general decision model can
identify favorable antenna combinations from numerical channel
information, and how much explicit wireless knowledge improves
its decisions.

\begin{figure}[t]
    \centering
    \includegraphics[width=\columnwidth]{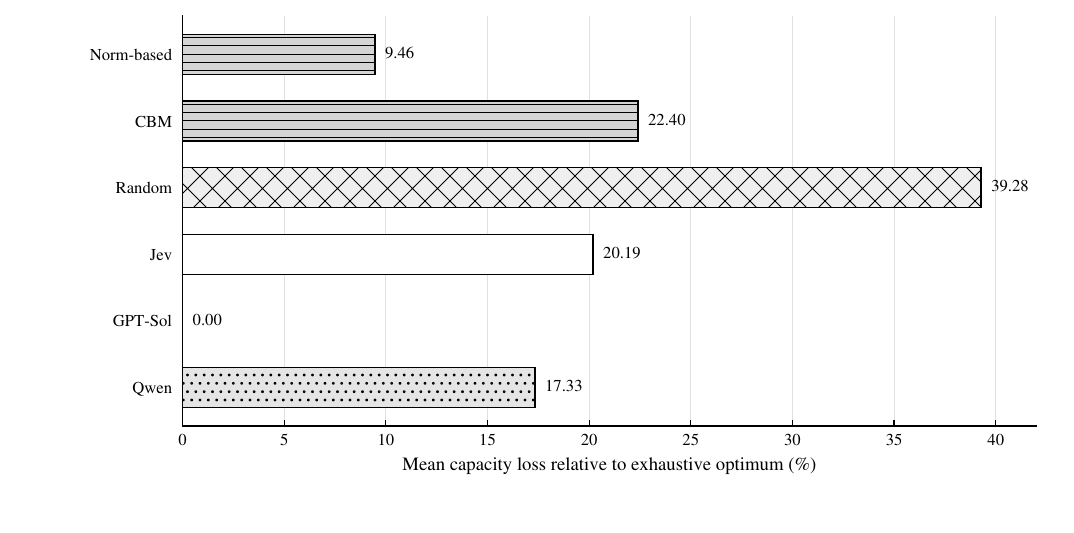}
    \caption{Mean relative capacity loss with respect to exhaustive
    search for receive-antenna selection.}
    \label{fig:antenna-quality}
\end{figure}

\begin{figure}[t]
    \centering
    \includegraphics[width=\columnwidth]{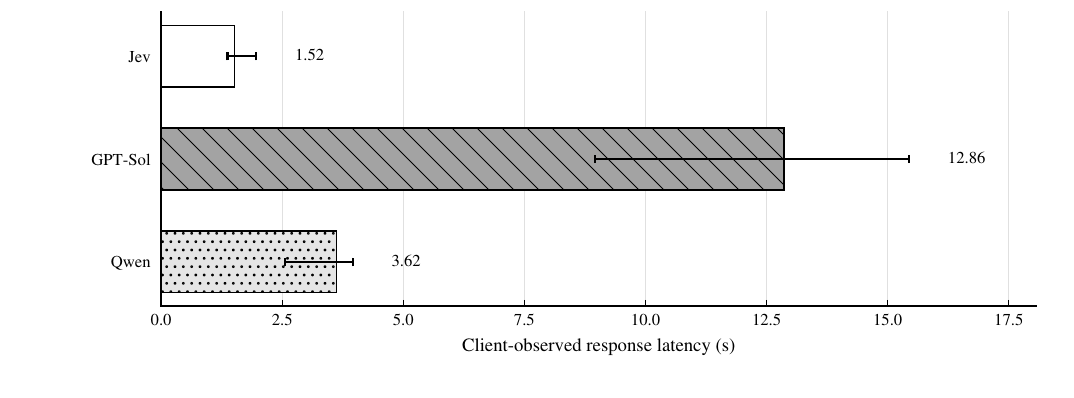}
    \caption{Client-observed response latency of the AI methods
    for receive-antenna selection.}
    \label{fig:antenna-latency}
\end{figure}

Figs.~\ref{fig:antenna-quality} and~\ref{fig:antenna-latency}
illustrate a tradeoff between decision quality and response latency.
Jev has a client-observed latency of 1.52~s, compared with
12.86~s for GPT-Sol and 3.62~s for Qwen, corresponding to
approximately $8.5\times$ and $2.4\times$ faster responses,
respectively. However, its mean relative capacity loss is
20.19\%, versus 0.00\% for GPT-Sol and 17.33\% for Qwen;
the simple norm-based baseline also achieves a lower loss of
9.46\%. Thus, in this mathematically demanding selection task,
Jev's faster response does not translate into better numerical
decisions than the evaluated LLMs. These findings suggest that
a low-latency, bounded decision interface alone is insufficient
for accurate channel-dependent optimization. Given the early
stage of System-One models, improving their numerical reasoning
and wireless-specific grounding remains an important direction,
potentially through integration with lightweight numerical
solvers. Nevertheless, the present results cover only one task
and should not be interpreted as a general limitation of the
model class.

\subsection{Scenario II: Intent-Conditioned RAN Slicing}

The second scenario allocates 100 physical resource blocks among
enhanced mobile broadband (eMBB), ultra-reliable low-latency
communications (URLLC), and massive machine-type communications
(mMTC). Allocations use increments of 20 resource blocks, producing
21 admissible actions. The benchmark comprises 30 instances formed
from five service-intent profiles and six network-state variants.

Each AI method receives a natural-language intent together with
traffic demand, channel conditions, and queue information. A
simplified numerical simulator scores the selected allocation using
service satisfaction and penalties for applicable service-level
agreement (SLA) violations. The underlying numerical intent weights
and SLA annotations are reserved for evaluation. Exhaustive
enumeration provides a full-information utility reference, while
load-aware and random allocation provide practical baselines.
The load-aware policy uses network state without interpreting
the textual intent.

The principal metrics are utility regret relative to the exhaustive
reference, SLA violation severity, the fraction of allocations
satisfying all applicable hard requirements, and decision latency.
Utility and constraint satisfaction are assessed separately because
the penalty-based utility optimum need not satisfy every hard
requirement. Overloaded instances also allow examination of
decisions when no admissible allocation satisfies all requirements.
This scenario evaluates the ability to translate service intent
into a bounded resource-allocation decision under competing demands.

\begin{figure}[t]
    \centering
    \includegraphics[width=\columnwidth]{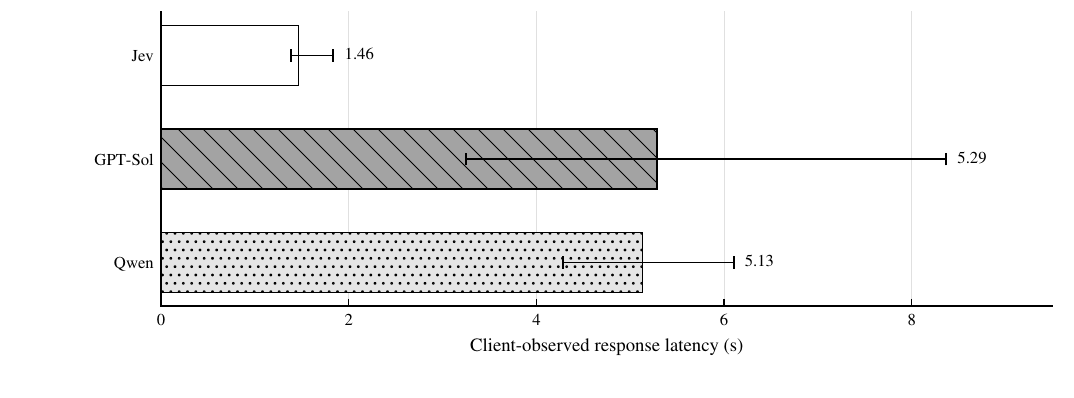}
    \caption{Client-observed response latency of the AI methods
    for intent-conditioned RAN slicing.}
    \label{fig:slicing-latency}
\end{figure}

\begin{figure}[t]
    \centering
    \includegraphics[width=\columnwidth]{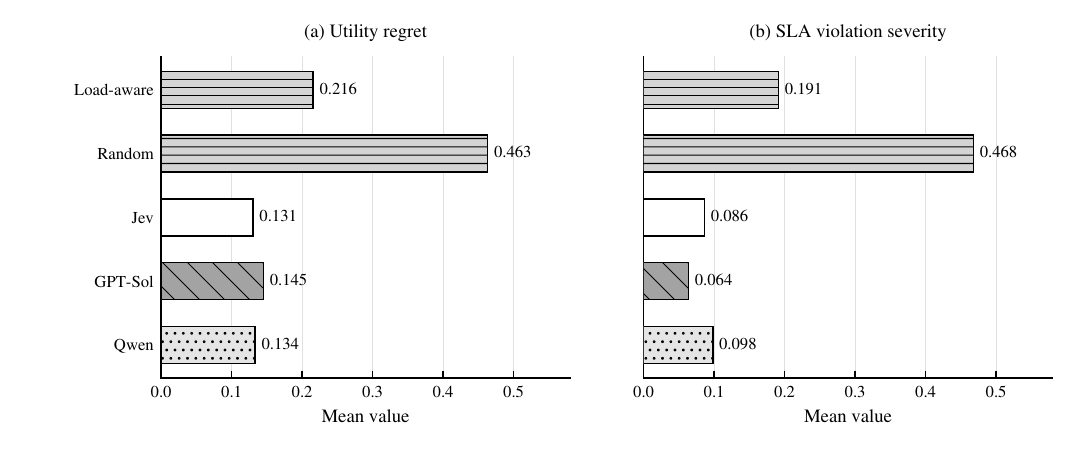}
    \caption{Decision quality for intent-conditioned RAN slicing.}
    \label{fig:slicing-quality}
\end{figure}

Figs.~\ref{fig:slicing-quality} and~\ref{fig:slicing-latency} show a different behavior from the numerically structured antenna-selection experiment. In the intent-conditioned RAN slicing task, the learned decision models clearly outperform the random and load-aware baselines in terms of decision quality. In particular, Jev achieves a mean utility regret of 0.131, which is comparable to Qwen (0.134) and GPT-Sol (0.145), while maintaining substantially lower inference latency. As shown in Fig.~\ref{fig:slicing-latency}, Jev reaches a median client-observed latency of only 1.46~s, compared with 5.13~s for Qwen and 5.29~s for GPT-Sol. These results suggest that, although current System-One models may still face limitations in tasks dominated by precise numerical computation, they can already provide competitive performance for compact intent-conditioned decision problems, where semantic requirements must be rapidly mapped to discrete control actions. Their favorable quality--latency tradeoff therefore makes them particularly promising for latency-sensitive wireless control as this emerging model family continues to mature.

\subsection{Scenario III: Intent-Driven Edge-Service Orchestration}

The third scenario draws on the independent evaluation by
Li et al.~\cite{li2026jev}. Jev translates natural-language
intents into service contracts specifying execution locations,
deadlines, and priorities, while a numerical scheduler assigns
tasks using current network and computing conditions.
The experiments combine live model responses, packet-level
New Radio simulation, mobility, shared edge queues, and
an actual image-reading service.

\begin{figure*}[t]
    \centering
    \includegraphics[width=\textwidth]{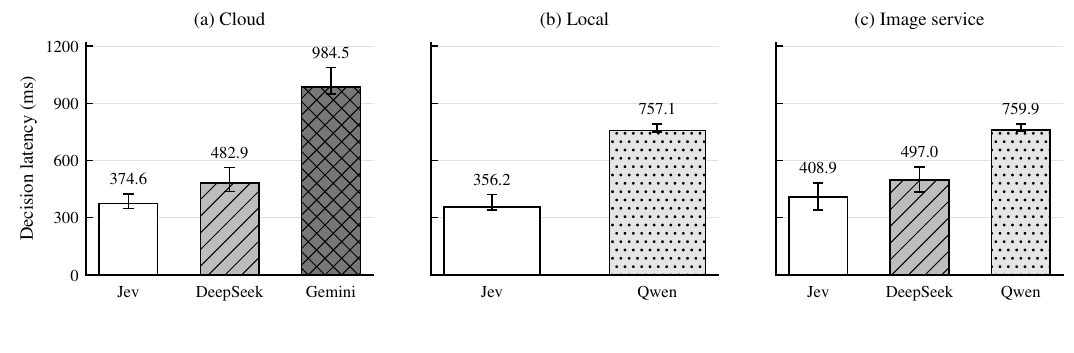}
    \caption{Decision latency for edge-service orchestration.
    Data from~\cite{li2026jev}.}
    \label{fig:edge-latency}
\end{figure*}

Fig.~\ref{fig:edge-latency} compares decision latency across
the cloud, local-model, and image-service experiments.
Bars indicate medians, and whiskers show the 25th--75th
percentiles rather than confidence intervals. Each experiment
uses separate measurements, including its own Jev measurements;
Qwen uses the direct-attribute interface throughout.
In the cloud comparison, Jev achieves a median latency of
374.6~ms, versus 482.9~ms for DeepSeek and 984.5~ms for Gemini.
Jev also provides lower median latency in the other two
settings. All compared interpreters achieve correct
interpretation on the evaluated contracts, with 63/63 checks
in the cloud and local-model experiments and 48/48 in the
image-service experiment.

\begin{figure*}[t]
    \centering
    \includegraphics[width=\textwidth]{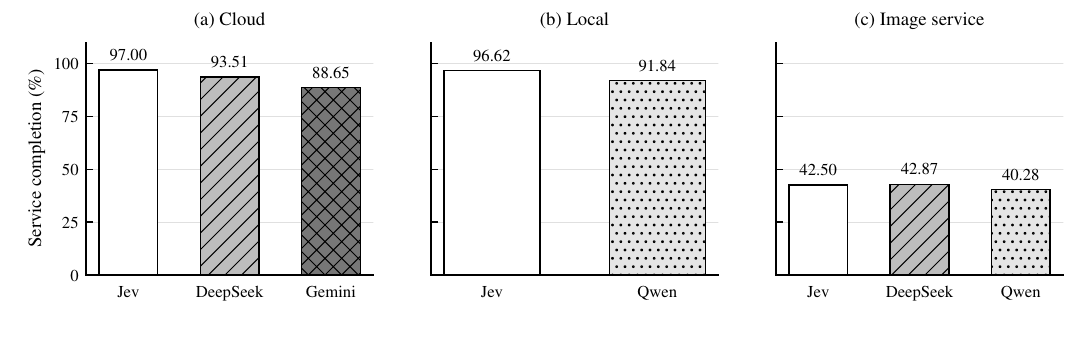}
    \caption{Service completion for edge-service orchestration.
    Data from~\cite{li2026jev}.}
    \label{fig:edge-completion}
\end{figure*}

Fig.~\ref{fig:edge-completion} presents the corresponding
service outcomes. Cloud and local-model results average
nine updated-contract trajectories, whereas image-service
results measure correct, on-time completion among 1,080
requests per method. These different workloads require
comparisons within each panel. In the cloud experiment,
completion reaches 97.00\% with Jev, compared with 93.51\%
for DeepSeek and 88.65\% for Gemini. The local-model
comparison likewise favors Jev, with 96.62\% completion
versus 91.84\% for Qwen. However, in the image service,
Jev completes 459 requests correctly and on time, compared
with 463 for DeepSeek and 435 for Qwen. Faster interpretation
therefore does not uniformly translate into higher
end-to-end completion.

These results illustrate the potential of combining bounded
intent interpretation with numerical scheduling, while
highlighting the need to assess application utility alongside
decision latency. Both figures are redrawn from Tables~2--4
of~\cite{li2026jev}; these externally reported results remain
separate from our simulations.

% =============================================================================
\section{Open Challenges and Research Directions}
\label{sec:challenges}
% =============================================================================

The preceding opportunities depend on capabilities that are not yet established
for System-One models in realistic wireless deployments. The main research
questions concern transferable decision learning, reliable uncertainty,
closed-loop safety, and deployability under network constraints.

\subsection{Wireless-Native Decision Foundation Models}

An important direction is a wireless-native System-One foundation model,
analogous in scope---but not in output mechanism---to foundation models in
language. Existing wireless foundation models learn reusable representations
from channel measurements, spectrograms, and heterogeneous wireless tasks
\cite{aboulfotouh2025radiofm,lwm2025,musefm2026,liang2026foundation}. A
decision-native model would instead be pretrained across many bounded control
problems. For tasks \(\tau\in\mathcal{T}\), its training data could be written as
\begin{equation}
    \mathcal{D}
    =\bigcup_{\tau\in\mathcal{T}}
    \{(\mathbf{s}^{\tau}_i,q^{\tau},
    \mathcal{A}^{\tau}_i,y^{\tau}_i)\}_{i=1}^{N_\tau},
\end{equation}
with the shared objective
\begin{equation}
    \min_\theta\sum_{\tau\in\mathcal{T}}w_\tau
    \mathbb{E}_{\mathcal{D}_\tau}
    \!\left[\ell_\tau\!\left(
    p_\theta(\cdot\mid\mathbf{s},q,\mathcal{A}),y\right)\right].
\end{equation}
The goal is to reuse wireless structure across PHY, RRM, mobility, and
management tasks while adapting to a new query or answer schema with little
task-specific data. Major obstacles include incompatible modalities and
timescales, inconsistent action semantics, scarce failure examples, negative
transfer, and the absence of large, representative decision--outcome datasets.

\subsection{Generalization and Calibrated Uncertainty}

Wireless distributions change with geography, frequency, hardware, mobility,
traffic, and network configuration. Consequently, strong results on random
train--test splits do not establish deployment robustness. Evaluation should
include held-out cells, environments, devices, and decision schemas, together
with out-of-distribution detection and risk--coverage analysis. Calibration
must also be maintained after adaptation or continual learning; otherwise, the
confidence used for abstention and escalation can become misleading precisely
when the environment changes.

\subsection{Constraint Satisfaction and Closed-Loop Safety}

A normalized decision distribution does not guarantee feasibility, stability,
or acceptable long-term behavior. Hard power, coverage, protocol, and security
constraints should therefore be enforced by deterministic checks or safety
shields. Temporal effects also require explicit treatment: individually
reasonable decisions may create ping-pong handovers, scheduler oscillations,
or repeated sleep/wake transitions. Combining System-One inference with
hysteresis, state history, optimizers, or digital twins is a promising route,
but the resulting hybrid system must be verified as a whole.

\subsection{Efficient Deployment and Reproducible Evaluation}

Practical performance is determined by end-to-end latency and energy, including
measurement collection, state construction, inference, validation, signaling,
and actuation. Compression, early exit, distributed inference, and
privacy-preserving adaptation should be studied together with deadline-aware
fallbacks. Progress also requires common datasets and benchmarks that compare
System-One models with heuristics, optimization, and task-specific learning at
matched compute and information budgets. In addition to average utility,
reporting should include constraint violations, tail latency, calibration,
selective risk, robustness under shift, and the frequency and cost of
escalation.

% =============================================================================
\section{Conclusion}
\label{sec:conclusion}

This paper examined System-One models as a decision-native alternative to autoregressive LLMs for wireless control problems with explicit and bounded action spaces.
Our evaluations show, however, that the value of this interface is strongly task dependent. 
In receive-antenna selection, Jev provided substantially lower response latency than the evaluated LLMs but could not match the decision quality of stronger numerical and LLM-based baselines.
In contrast, for intent-conditioned RAN slicing, Jev achieved decision quality comparable to the evaluated LLMs while reducing response latency by more than threefold.

Taken together, these results suggest that System-One models are most compelling
as a complementary inference layer for repeated, bounded, and latency-sensitive decisions, particularly when semantic intent must be mapped rapidly to discrete control actions. 
A promising wireless architecture is therefore hybrid: System-One models can handle routine high-confidence decisions, while 
more capable reasoning models address uncertain, constrained, or computationally demanding cases. 
Realizing this vision will require wireless-native training, reliable calibration under distribution shift, explicit constraint enforcement, and reproducible end-to-end evaluation. 
As these capabilities mature, decision-native foundation models may provide a useful building block for scalable and responsive intelligence in future wireless networks.
% =============================================================================

\bibliographystyle{IEEEtran}
\bibliography{references}

\end{document}